\documentclass[10pt,conference,letterpaper]{IEEEtran}
\pdfoutput=1 
\IEEEoverridecommandlockouts
\usepackage[utf8]{inputenc}
\usepackage[english]{babel}
\usepackage{multirow} 
\usepackage{booktabs}
\usepackage{threeparttable} 
\usepackage{cite}
\usepackage{amsmath,amssymb,amsfonts}
\usepackage{algorithmic}
\usepackage{graphicx}
\usepackage{textcomp}
\usepackage{xcolor}
\usepackage{xcolor}
\definecolor{tkPink}{HTML}{EA0A8E}
\definecolor{tkGray1}{HTML}{999999}
\definecolor{tkGray2}{HTML}{666666}
\definecolor{darkpastelgreen}{rgb}{0.01, 0.75, 0.24}
\definecolor{BCUred}{rgb}{0.8, 0.0, 0.0}
\definecolor{UOSred}{RGB}{172, 6, 52}
\definecolor{UOSgray}{RGB}{207, 207, 207}
\definecolor{verylightgray}{RGB}{242, 242, 242}
\def\BibTeX{{\rm B\kern-.05em{\sc i\kern-.025em b}\kern-.08em
    T\kern-.1667em\lower.7ex\hbox{E}\kern-.125emX}}

\usepackage{balance}
\usepackage[ruled, lined, linesnumbered, commentsnumbered, noend]{algorithm2e}

\usepackage{subcaption}
\usepackage{newfloat}
\DeclareFloatingEnvironment[fileext=frm,placement={tbp},name=Problem]{problem}
\usepackage{tikz}
\usetikzlibrary{spy,%
				arrows,%
                shapes,
                positioning,
                calc,
                automata,
                decorations.pathreplacing,
                angles,
                quotes,
                backgrounds,
                shadings,
                cd}
\tikzset{
    double color fill/.code 2 args={
        \pgfdeclareverticalshading[%
            tikz@axis@top,tikz@axis@middle,tikz@axis@bottom%
        ]{diagonalfill}{100bp}{%
            color(0bp)=(tikz@axis@bottom);
            color(50bp)=(tikz@axis@bottom);
            color(50bp)=(tikz@axis@middle);
            color(50bp)=(tikz@axis@top);
            color(100bp)=(tikz@axis@top)
        }
        \tikzset{shade, left color=#1, right color=#2, shading=diagonalfill}
    }
}

\tikzset{
  node split radius/.initial=1,
  node split color 1/.initial=red,
  node split color 2/.initial=green,
  node split color 3/.initial=blue,
  node split half/.style={node split={#1,#1+180}},
  node split/.style args={#1,#2}{
    path picture={
      \tikzset{
        x=($(path picture bounding box.east)-(path picture bounding box.center)$),
        y=($(path picture bounding box.north)-(path picture bounding box.center)$),
        radius=\pgfkeysvalueof{/tikz/node split radius}}
      \foreach \ang[count=\iAng, remember=\ang as \prevAng (initially #1)] in {#2,360+#1}
        \fill[line join=round, fill=\pgfkeysvalueof{/tikz/node split color \iAng}]
          (path picture bounding box.center)
          --++(\prevAng:\pgfkeysvalueof{/tikz/node split radius})
          arc[start angle=\prevAng, end angle=\ang] --cycle;
} } }

\graphicspath{{./figures/}}

\usepackage[nolist]{acronym}
\usepackage{hyperref}
\hypersetup{nolinks=true} 

\usepackage[absolute]{textpos} 

\begin{document}

\title{Preprocess your Paths -- Speeding up\\Linear Programming-based Optimization for\\Segment Routing Traffic Engineering}

\def\IEEEauthorrefmark#1{\raisebox{0pt}[0pt][0pt]{\textsuperscript{\footnotesize\ensuremath{\ifcase#1\or *\or \dagger\or \ddagger\or%
    \bullet\or \circ\or \cdot\or \times\or \checkmark\or **\or \dagger\dagger%
    \or \ddagger\ddagger \else\textsuperscript{\expandafter\romannumeral#1}\fi}}}}

\author{
  \IEEEauthorblockN{Alexander Brundiers\IEEEauthorrefmark{5}, Timmy Sch\"{u}ller\IEEEauthorrefmark{4}\IEEEauthorrefmark{5}, Nils Aschenbruck\IEEEauthorrefmark{5}}
\vspace*{.18cm}
\IEEEauthorblockA{
  \begin{tabular}{ccc}
    {\IEEEauthorrefmark{5}Osnabr\"{u}ck University, Institute of Computer Science, Germany} && {\IEEEauthorrefmark{4}Deutsche Telekom Technik GmbH, Germany}\\
    {Email: \{brundiers, schueller, aschenbruck\}@uos.de} && {Email: timmy.schueller@telekom.de}\\
  \end{tabular}
}}

\maketitle


\acrodefplural{PoP}[PoPs]{Points of Presence}
\begin{acronym}
\acro{DEFO}{Declarative and Expressive Forwarding Optimizer}
\acro{DP}{Demand Pinning}
\acro{ECMP}{Equal Cost Multipath}
\acro{E2E}{end-to-end}
\acro{GSP}{Group Shortest-Path}
\acro{H2SR}{Hybrid 2SR}
\acro{HSLS}{Hybrid SR LS}
\acro{IGP}{Interior Gateway Protocol}
\acro{IE}{Ingress-Egress}
\acro{ISP}{Internet Service Provider}
\acro{IS-IS}{Intermediate System to Intermediate System}
\acro{LDP}{Label Distribution Protocol}
\acro{LER}{Label Edge Router}
\acro{LP}{Linear Program}
\acro{LS}{Local Search}
\acro{LSR}{Label Switched Router}
\acro{LSP}{Label Switched Path}
\acro{MCF}{Multicommodity Flow}
\acro{MLU}{Maximum Link Utilization}
\acro{MO}{Midpoint Optimization}
\acro{MOLS}{Midpoint Optimization Local Search}
\acro{MPLS}{Multiprotocol Label Switching}
\acro{MSD}{Maximum Segment Depth}
\acro{NDA}{non-disclosure agreement}
\acro{OSPF}{Open Shortest Path First}
\acro{PD}{Path Domination}
\acro{PoP}{Point of Presence}
\acro{RSVP}{Resource Reservation Protocol}
\acro{RL2TLE}{Router-Level 2TLE}
\acro{SB}{Stretch-Bounding}
\acro{SC2SR}{Shortcut 2SR}
\acro{SID}{Segment Identifier}
\acro{SPR}{Shortest Path Routing}
\acro{SR}{Segment Routing}
\acro{SRLS}{Segment Routing Local Search}
\acro{TE}{Traffic Engineering}
\acro{TLE}{Tunnel Limit Extension}
\acro{WAE}{WAN Automation Engine}
\acro{w2TLE}{Weighted 2TLE}
\end{acronym}

\begin{abstract}
Many state-of-the-art \ac{SR} \ac{TE} algorithms rely on \ac{LP}-based optimization.
However, the poor scalability of the latter and the resulting high computation times impose severe restrictions on the practical usability of such approaches for many use cases.
To tackle this problem, a variety of preprocessing approaches have been proposed that aim to reduce computational complexity by preemtively limiting the number of \ac{SR} paths to consider during optimization.
In this paper, we provide the first extensive literature review of existing preprocessing approaches for \ac{SR}.
Based on this, we conduct a large scale comparative study using various real-world topologies, including recent data from a Tier-1 \ac{ISP} backbone.
Based on the insights obtained from this evaluation, we finally propose a combination of multiple preprocessing approaches and show that this can reliably reduce computation times by around a factor of 10 or more, without resulting in relevant deterioration of the solution quality.
This is a major improvement over the current state-of-the-art and facilitates the reliable usability of \ac{LP}-based optimization for large segment-routed networks.
\end{abstract}

\acresetall


\section{Introduction}
\ac{SR} has become a premier choice for \ac{TE} purposes for large networks.
It offers great traffic steering capabilities while simultaneously offering good scalability.
However, in order to use \ac{SR} to its full potential, optimization algorithms are needed to compute the best possible \ac{TE} configurations.
In many state-of-the-art approaches (e.g., \cite{bhatia}, \cite{midpoint_optimization_infocom}, or \cite{schueller_ton}), this is done using \ac{LP}-based optimization because it can provide guaranteed optimal solutions.
Its major drawback, however, is its limited scalability and the resulting high computation times for larger networks.
Depending on the algorithm and the size of the network, computation times can reach up to multiple hours or even days.
For certain use cases, this is acceptable but for many scenarios, such high computation times severely limit the practical usability of \ac{LP}-based \ac{SR} \ac{TE} algorithms.
\par
Over the recent years, a variety of preprocessing approaches have been proposed that aim to reduce the problem complexity and, thus, the resulting computation time, by preemptively limiting the number of \ac{SR} paths to consider during optimization.
While the individually reported results for those approaches look promising, evaluations are often carried out on a rather limited set of data and varying hardware.
This raises questions regarding the generalizability of the results and makes it virtually impossible (i.e. for operators) to compare approaches against each other to select the best fitting one. 
\par
To address these problems, we provide an extensive literature review and discussion of existing preprocessing approaches to then carry out a large comparative study regarding their performance.
For this, we not only use various publicly available topologies from the Repetita dataset \cite{repetita} but also recent network data from the backbone of a globally operating Tier-1 \ac{ISP}.
Finally, based on insights gained from this evaluation, we propose a combination of multiple preprocessing approaches and show that this leads to a significant improvement in performance.
It allows for a reduction of computation times by a factor of 10 or more reliably without a relevant deterioration in solution quality.
This is a major improvement over the current state-of-the-art and an important step towards the reliable usability of \ac{LP}-based \ac{SR} \ac{TE} for large networks.

\section{An Introduction to Segment Routing} \label{sec:background}
\ac{SR} \cite{srArch} is a network tunneling technique that implements the source routing paradigm.
Its key feature is the possibility to add specific labels (also called \textit{segments}) to a packet, which function as waypoints that the packet has to visit in a given order before heading to its original destination.
The forwarding paths between the waypoints are determined by the \ac{IGP} of the respective network.
Overall, \ac{SR} enables the definition of virtually arbitrary forwarding paths and allows for a precise, per-flow traffic control.
For this reason, \ac{SR} has become one of the premier choices for \ac{TE} and there is a large body of work regarding \ac{SR} in general and its applications for \ac{TE} in particular (cf. e.g., \cite{srSurvey}).
\par
One of the fundamental works in the \ac{SR} \ac{TE} landscape is \cite{bhatia}.
Here, the authors propose an \ac{LP}-based optimization model that builds the foundation for many subsequent works.
A slightly adapted version of the respective \ac{LP} formulation is shown in Problem \ref{problem:2sr_lp}.
The objective is to minimize the \ac{MLU} denoted by $\theta$.
The variables $x_{ij}^k$ indicate the percentage share of the demand $t_{ij}$ between nodes $i$ and $j$, that is routed over the intermediate segment $k$.
Equation (\ref{eq:2sr_1}) ensures that each demand is satisfied.
Equation (\ref{eq:2sr_2}), together with the objective function, minimizes the \ac{MLU}. 
For every edge $e$, $g_{ij}^k(e)$ indicates the load that is put on $e$ if a uniform demand is routed from $i$ to $j$ over the intermediate segment $k$.
These values are constants and can be efficiently precomputed.
All in all, the left side of the constraint denotes the traffic that is put on $e$ by the \ac{SR} configuration represented by the $x_{ij}^k$.
This is then limited to the edges capacity $c(e)$ scaled by $\theta$.
By minimizing this scaling factor, a \ac{SR} configuration with minimal \ac{MLU} is computed.
The only difference to the original \ac{LP} of \cite{bhatia} is that there the $x_{ij}^k$ variables were continuous, allowing for demands to be split arbitrarily across various \ac{SR} paths.
However, such an arbitrary splitting is not feasible in practice \cite{schueller_ton}. Therefore, newer variations of the 2SR \ac{LP} generally prohibit splitting demands over multiple \ac{SR} paths by making the $x_{ij}^k$ binary variables (cf. e.g., \cite{sr_preprocessing} or \cite{schueller_ton}).

\begin{problem}
\centering
{\small
\begin{minipage}{0.85\linewidth}
    \begin{flalign}
     & \makebox[0pt][l]{$\displaystyle{}\text{min } \theta$} \label{eq:2sr_obj}\\
     & \text{s.t.} & \sum_{k}{x^{k}_{ij}}   &\; = \;  1 && \forall ij && \label{eq:2sr_1}\\
     &             & \sum_{ij}{t_{ij}\sum_{k}{g_{ij}^{k}(e\,)x_{ij}^{k}}}         & \; \leq \; \theta \, c(e)                  && \forall e        && \label{eq:2sr_2} \\ 
     &             & x_{ij}^{k} & \; \in \; \{0,1\} && \forall ijk
    \end{flalign}
\end{minipage}
}
\medskip
\caption{2\acs{SR} formulation (inspired by \cite{bhatia}).}\label{problem:2sr_lp}
\end{problem}

\par
A recent innovation in the \ac{SR} landscape is the \textit{\ac{MO}} concept \cite{midpoint_optimization_infocom}.
Demands no longer have to be optimized individually by deploying dedicated end-to-end \ac{SR} tunnels. Instead, a single \ac{SR} tunnel can be used to detour a whole set of demands.
\ac{MO} allows for a substantial reduction of the number of \ac{SR} tunnels that need to be deployed to implement \ac{TE} solutions, lowering the configuration effort and overhead in the network.
However, the underlying optimization problem becomes inherently more complex, resulting in substantially higher computation times.


\section{Preprocessing Approaches for LP-based Segment Routing Optimization} \label{sec:sr_middlepoint_selection}
A major challenge in the context of \ac{SR} \ac{TE} is the scalability of the used optimization algorithms.
While \ac{LP}-based approaches offer the major advantage of providing provable optimal solutions, they scale rather poorly with network size.
For small to medium sized networks, this is no issue since solutions can still be computed within seconds or at most minutes.
However, for large networks (e.g., WANs or ISP backbones), computing \ac{TE} solutions with \acp{LP} can take multiple hours or more (cf. \cite{schueller_ton}), while, in practice, solutions might be needed on a timescale of just a few minutes (cf. \cite{defo2, microsoft_swan}).
There are different ways to approach these issues.
Some focus on the use of advanced mathematical concepts like \textit{column generation} \cite{cg4sr} or \textit{constraint programming} \cite{defo2}, while others try to deploy meta-heuristics to compute reasonable good solutions within really short timespans (e.g., \cite{mols_heuristic} or \cite{srls}).
\par
A completely different approach to bring down the complexity and, hence, computation time of \ac{SR} \acp{LP} focuses on \textit{preprocessing} the set of \ac{SR} paths to consider during optimization.
Each \ac{SR} path basically consists of the source and destination node of the packet as well as a set of \textit{middlepoints}\footnote{The term ``\textit{middlepoint}'' used in the remainder of this paper is \textbf{not} to be confused with the term ``\textit{midpoint}'' from the \ac{MO} concept (Section \ref{sec:background}).} (the node segments) that it has to visit (cf. Section \ref{sec:background}).
In basically all \ac{SR} \ac{LP} formulations (e.g., in Problem \ref{problem:2sr_lp}), the model allows for every node segment to be used as a middlepoint for each traffic demand (aka. source-destination pair).
While this guarantees optimality, it also is responsible for a large portion of the overall problem complexity.
For every demand the optimization has $|V|^{k-1}$ paths to choose from, resulting in a total number of $\mathcal{O}(|V|^{k+1})$ possible \ac{SR} paths to evaluate, with $|V|$ being the number of nodes in the network and $k$ the maximum number of segments per path.
Here, the preprocessing (or also called \textit{middlepoint selection} \cite{centrality_based_middlepoint_selection}) approaches come into play and try to reduce this complexity by limiting the set of available middlepoints per demand and, thus, the set of \ac{SR} paths to consider prior to optimization.
This results in smaller and generally faster to solve \acp{LP}.
\par
In the following, we provide a detailed overview on (to the best of our knowledge) all existing preprocessing approaches in the \ac{SR} \ac{TE} literature.

\subsection{Centrality-based Approaches} \label{subsec:centrality_based}
One of the first works that came up with the idea of preemptively limiting the number \ac{SR} paths that are considered for optimization are \cite{node_constrained_te_ton, centrality_based_middlepoint_selection}.
They only allow a certain subset of nodes as \textit{middlepoints} (aka. intermediate segments) for \ac{SR} paths and propose to use \textit{graph centrality} metrics to select ``important'' or ``central'' nodes into this subset.
This idea is evaluated in the context of datacenter networks and \ac{ISP} backbones for various subset sizes and with different centrality measures.
It is observed that, out of all considered centrality metrics, selecting the allowed middlepoints based on their \textit{\ac{GSP} centrality} \cite{gsp_centrality} performs best. 
For a group of nodes $\mathcal{G}$ the \ac{GSP} centrality is defined as:
\begin{equation}
C_{gsp}(\mathcal{G}) =\sum_{s,t\in V | s,t \not \in \mathcal{G}} \frac{\theta_{st}(\mathcal{G})}{\theta_{st}}
\end{equation}
with $\theta_{st}$ denoting the total number of shortest paths from $s$ to $t$ and $\theta_{st}(\mathcal{G})$ being the number of shortest paths from $s$ to $t$ that include any node in $\mathcal{G}$ \cite{centrality_based_middlepoint_selection}.
In other words, it characterizes how ``central'' a group of nodes is based on the number of shortest paths that run through this group.
\par
Overall, it is shown that, when focusing on only a small number of nodes as available middlepoints, computation times can be substantially reduced.
However, this comes at the prices of a considerable deterioration in solution quality.
While the authors argue that this can be a sensible trade-off to make, in practice, a deterioration of solution quality is only acceptable up to a certain point.
Furthermore, limiting the available middlepoints to the same small set of nodes for all demands can result in severe violations of certain operational latency constraints (i.e. from service level agreements).
For example, if, for a globe-spanning network, the most ``central'' nodes are all located in Europe, intra-US traffic either needs to always follow its shortest path or be detoured all the way over a node in Europe, most likely exceeding latency bounds. 
In addition to that, if the number of \ac{SR} paths grows larger and they are all forced over the same handful of middlepoints, this can put additional burden on the routing hardware of these nodes.

\subsection{Stretch-Bounding}\label{subsec:stretch_bounding}
Another early middlepoint selection approach is the \textit{\ac{SB}} concept proposed in \cite{stretch_bounding_1, stretch_bounding_2}.
Its key idea is to rule out all nodes from being a potential middlepoint for an \ac{SR} path if they are ``too far away'' from its source or destination regarding a given metric.
This can be formalized as only considering middlepoint $m$ for an \ac{SR} path between $src$ and $dst$ if the following equation is satisfied:
\begin{equation}
\frac{\mathit{DIST}(src\rightarrow m) + \mathit{DIST}(m\rightarrow dst)}{\mathit{DIST}(src\rightarrow dst)} \leq \alpha_{SB}
\label{eq:stretch_bounding}
\end{equation}
with the $\mathit{DIST}()$ function denoting the shortest path distance between the respective two nodes and $\alpha_{SB} \in [1,\infty]$ being the so called \textit{\ac{SB} factor}.
This approach rules out all those \ac{SR} paths that are more then $\alpha$-times longer than the shortest path between the respective source and destination.
It is shown in \cite{stretch_bounding_2} that there is a trade-off between speedup and deterioration of solution quality depending on the chosen $\alpha$-value.
The authors define a factor of around $\alpha_{SB} = 1.4$ as the sweetspot of achieving close to optimal results while still considerably speeding up computations by a factor of 3 to 4.
However, the \ac{SB} implementation as described in Equation \ref{eq:stretch_bounding} inherits an issue that can negatively impact performance in certain scenarios (first pointed out in \cite{sr_preprocessing}).
If the initial shortest path length is small (e.g., for paths with just one or two hops), small $\alpha$-values can completely prohibit any kind of detour for the respective demand.
The best example for this is a simple hop-count metric. If the shortest path of a demand has length 1 (one hop), this means that for all $\alpha < 2$, there are no detours available for this demand as every detour would have at least length two. This can negatively impact the achievable \ac{MLU}.
\par
A rather similar concept to the \ac{SB} approach of \cite{stretch_bounding_1} was also proposed in \cite{schueller_ton_2}, where nodes are assigned geographical tags on three different levels of granularity (\texttt{site}, \texttt{country}, and \texttt{continent}).
\ac{SR} paths between nodes that share a common tag value (e.g., \textit{US} for the \texttt{country}-tag) are restricted to only use middlepoints with the same tag value (e.g., only nodes also located in the US).
This also implements the idea of limiting the length of \ac{SR} detours to a sensible maximum (e.g., by not routing traffic between Boston and New York over Europe).

\subsection{Demand Pinning}\label{subsec:demand_pinning}
Another preprocessing technique briefly described by \textit{Microsoft} \cite{demand_pinning_1, demand_pinning_2} is \textit{\ac{DP}}.
It is based on the observation that in many networks (i.e. WANs and \ac{ISP} backbones) traffic flows are not uniformly distributed in size.
Instead, traffic consist of a few very large demands that make up a considerable amount of the total traffic volume and a rather large number of very small demands.
\ac{DP} fixes the forwarding paths of all these small demands to standard \ac{SPR} and only runs a \ac{TE} optimization for the larger ones.
The idea is that the impact of the small demands on the overall solution quality is negligible compared to the larger traffic flows. Not optimizing them will have virtually no impact on the overall solution quality.

\subsection{SR Path Domination}\label{subsec:policy_domination}
All of the previous approaches carry the risk of excluding \ac{SR} paths that are needed for an optimal solution.
As a result, the solution quality can become arbitrarily worse when deploying these methods.
To prevent this, one has to ensure to only exclude \ac{SR} paths for which it can be proven that they are not needed for an optimal solution.
A first step towards such an approach was presented in \cite{weak_loops}. There, it is shown that a large portion of configurable SR paths actually contain loop-like structures and the authors suspect that many of these paths are not required to obtain optimal solutions.
This assumption is further investigated and confirmed by Callebaut et al. \cite{sr_preprocessing}.
They propose the concept of \textit{dominated} and \textit{equivalent} \ac{SR} paths.
An \ac{SR} path $p_1$ is \textit{dominated} by another path $p_2$ if three conditions are satisfied.
First, both paths must have the same start- and endpoint.
Second, assuming a uniform traffic flow is routed over each path, for each link $l$ in the set $\mathcal{L}(p_2)$ of links used by $p_2$ the load put on $l$ by $p_2$ must be lower or equal to the load put on $l$ by $p_1$:
\begin{equation}
 load(l, p_2) \leqslant load(l, p_1) \qquad \forall \, l \in \mathcal{L}(p_2) \label{eq:domination_3}
\end{equation}
Lastly, for at least one link in Equation \ref{eq:domination_3} the strict inequality must hold.
Analogously, two \ac{SR} paths are \textit{equivalent}, if their set of used links and the resulting link-loads are exactly equal. This is the case if the first two conditions for \ac{SR} path domination hold but with exact equality for Equation \ref{eq:domination_3}.
\par
Dominated \ac{SR} paths are never needed for an optimal solution and for a set of equivalent paths, it is sufficient to consider just one of them, allowing to exclude all others.
It is shown in \cite{sr_preprocessing} that, based on these two observations, a substantial number of \ac{SR} paths can be ruled out prior to optimization, resulting in a significant reduction in computation time.
Furthermore, just like \ac{SB} and centrality-based preprocessing approaches, this just requires information on the network topology but not on traffic. Hence, it can be precomputed in advance which is quite useful since in \cite{sr_preprocessing}, computation times of up to 30min or more are reported for just the preprocessing of larger topologies.

\subsection{Discussion} \label{subsec:rel_work_discussion}
As described in the previous sections, there is a wide variety of possible preprocessing approaches for \ac{SR}.
However, judging and comparing the quality and usefulness of these different approaches proves to be difficult.
The reasons for this are manifold:
Meaningful cross comparisons between the publications are virtually impossible as they all use 1) different hardware as well as 2) varying datasets.
Furthermore, 3) basically all evaluations are carried out on (semi-)artificial data, like the Repetita dataset.
Even if real-world data is used (e.g., from the \textit{Geant} network in \cite{stretch_bounding_2}), it is rather old and mostly from research networks which do not feature the same characteristics as large \ac{ISP} backbones.
Hence, it is unclear whether the results obtained on such data are directly transferable to a practical application in large commercial networks.
4) While some works (e.g., \cite{sr_preprocessing}) feature an extensive evaluation on a large set of different networks, others (e.g., \cite{stretch_bounding_2} or \cite{centrality_based_middlepoint_selection}) only test their approaches on a very limited number of networks (6 and 2, respectively).
Even though their results look promising, the sample size is probably far to low to allow for a meaningful generalization of the results to other networks.
And lastly, 5) while some approaches (e.g., \ac{DP}) sound very promising in theory, there are no evaluation results reported in the literature, at all.
\par
We aim to address the above issues by carrying out an extensive performance evaluation of all preprocessing approaches on a large set of networks from the Repetita dataset as well as recent network data from a globally operating Tier-1 \ac{ISP}.

\section{Evaluation Setup} \label{sec:evaluation_setup}
This section presents our evaluation setup by introducing the used datasets and describing the respective algorithm implementations.
All computations are carried out on the same 64-core 3.3GHz machine with around 500GB of RAM and using CPLEX 20.1.0 \cite{cplex} as LP-solver.

\subsection{Data}
We carry out our evaluation on two sets of data.
The first one consists of data from the publicly available \textit{Repetita} dataset \cite{repetita}.
It features topologies of real-world networks (mostly WANs or \ac{ISP} backbones) collected in the \textit{Internet Topology Zoo} \cite{topo-zoo} and artificially generated traffic matrices (using a \textit{random gravity model} \cite{traffic_matrix_synthesis}) for each topology.
In addition to that, each topology also comes with two sets of \ac{IGP} metrics (\textit{unary} and \textit{inverse capacity}).
However, we limit our evaluations to only the \textit{unary} metric set as previous results \cite{schueller_metrics} have shown that the impact of different metric designs on \ac{SR} performance is negligible. Other results indicate that \ac{SR} middlepoint selection approaches also seem to be quite robust regarding the underlying metric (cf. e.g., \cite{stretch_bounding_2}).
We also discard all those instances which are already solved optimally by \ac{SPR} since for those, there is no optimization to be done.
Finally, since for smaller networks with just a couple tens of nodes, even rather complex \acp{LP} are generally solvable within seconds or less (cf. e.g., \cite{hybrid_sr}), there is basically no practically relevant improvement to achieve for these networks.
Therefore, we limit our evaluations to larger networks with at least 50 nodes.
This leaves us with a total of 57 networks comprising of 50 to 197 nodes and around 120 to 500 edges.
\par
Complementary to the Repetita data with artificial traffic, we also carry out evaluations on a second set of data collected from the backbone network of a globally operation Tier-1 \ac{ISP}.
It features 18 topology snapshots that resemble different expansion states of the network between 2017 and 2021 and a real traffic-matrix collected during the peak-hour of the respective day.
Depending on the expansion state of the network, the topology features around 100 to 200 nodes and 600 to 1100 edges. 

\subsection{Algorithms \& Implementations}
We limit our evaluation of the centrality-based middlepoint selection approaches to the \ac{GSP} centrality as it was identified as performing best in previous works (cf. Section \ref{subsec:centrality_based}).
To get around the issues regarding the high algorithmic complexity of its computation \cite{centrality_based_middlepoint_selection} (and the resulting high computation times), we use an approximation algorithm provided by \textit{NetworKit}\footnote{https://networkit.github.io/} which approximates the node group with maximum centrality up to a given accuracy $\epsilon$.
Such an approximation would (most likely) also be used in a practical deployment due to the substantial performance gains.
For our evaluations, we use $\epsilon = 0.005$ which allows to compute the respective maximum centrality group in a couple seconds for most instances.
We also cross-validated the approximation results against those of an exact algorithm for some of the smaller instances and the results matched virtually perfectly.
We implement the \ac{SB} approach as described in Section \ref{subsec:stretch_bounding}, with a small extension to address the already described issues regarding demands with very low shortest path lengths.
For this, we allow each demand with a shortest path length of just one hop to be rerouted over arbitrary paths with two hops (irrespective of the chosen $\alpha$-value).
This turns out to be sufficient to resolve most of the respective issues without significantly increasing the overall number of \ac{SR} paths to consider during optimization.
The same was also observed in \cite{sr_preprocessing}.
To implement \ac{DP}, we first sort all traffic demands by size in ascending order.
After this, we keep fixing the smallest demands to their \ac{SPR} paths until the total sum of ``fixed'' traffic reaches a certain share of the total traffic volume given by the parameter $\alpha_{DP} \in [0,1]$.
\par
Similar to the related work, we evaluate the effectiveness of the preprocessing approaches based on the 2SR \ac{LP} (Problem~\ref{problem:2sr_lp}). It is the de-facto standard \ac{LP} for \ac{SR} \ac{TE} and builds the foundations for a wide body of derivative work (cf. Section \ref{sec:background}) to which the findings should be transferable.
Our evaluation focuses on the resulting \ac{MLU} deterioration and the achievable speedup compared to the standard 2SR implementation.

\section{Evaluation Results} \label{sec:evaluation_results}
\begin{figure*}
     \centering
     \begin{subfigure}[b]{0.32\textwidth}
         \centering
         \includegraphics[width=0.95\textwidth]{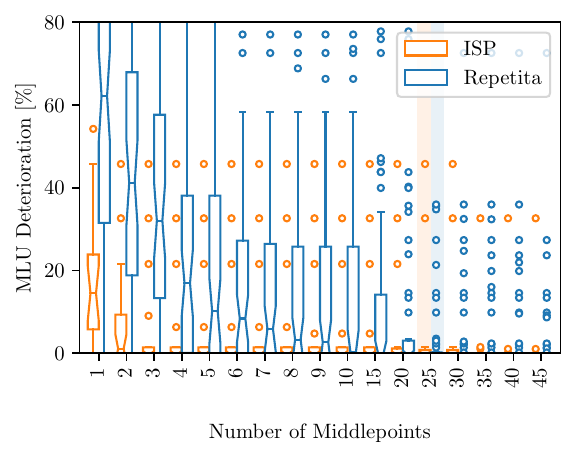}
         \caption{Centrality-based}
         \label{subfig:centrality_mlu}
     \end{subfigure}
     \hfill
     \begin{subfigure}[b]{0.32\textwidth}
         \centering
         \includegraphics[width=0.96\textwidth]{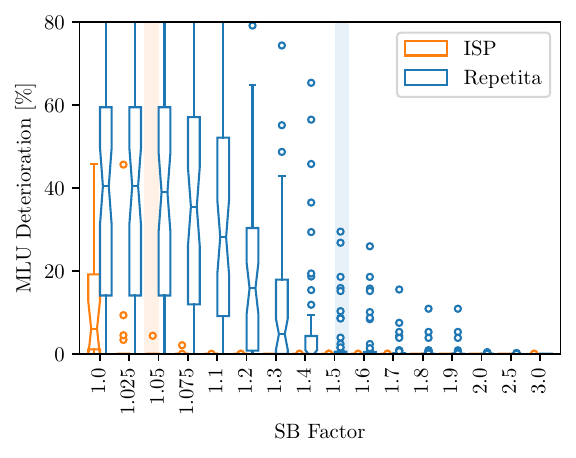}
         \caption{Stretch Bounding (extended)}
         \label{subfig:stretchbounding_mlu}
     \end{subfigure}
     \hfill
     \begin{subfigure}[b]{0.32\textwidth}
         \centering
         \includegraphics[width=0.96\textwidth]{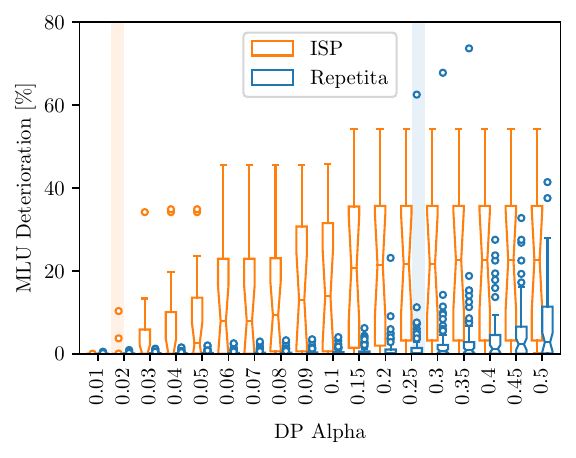}
         \caption{Demand Pinning}
         \label{subfig:deman_pinning_mlu}
     \end{subfigure}
        \caption{MLU deterioration for different preprocessing approaches. (A few very large outliers were cut off for better readability.)}
        \label{fig:approach_comparison_mlu}
\end{figure*}

\begin{figure*}
     \centering
     \begin{subfigure}[b]{0.32\textwidth}
         \centering
         \includegraphics[width=0.96\textwidth]{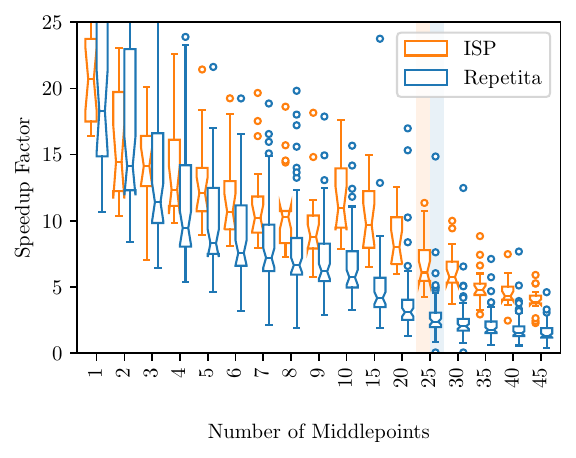}
         \caption{Centrality-based}
         \label{subfig:centrality_speedup}
     \end{subfigure}
     \hfill
     \begin{subfigure}[b]{0.32\textwidth}
         \centering
         \includegraphics[width=0.96\textwidth]{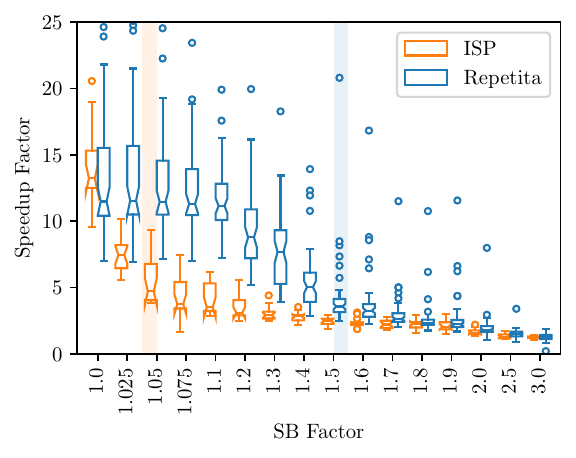}
         \caption{Stretch Bounding (extended)}
         \label{subfig:stretchbounding_speedup}
     \end{subfigure}
     \hfill
     \begin{subfigure}[b]{0.32\textwidth}
         \centering
         \includegraphics[width=0.96\textwidth]{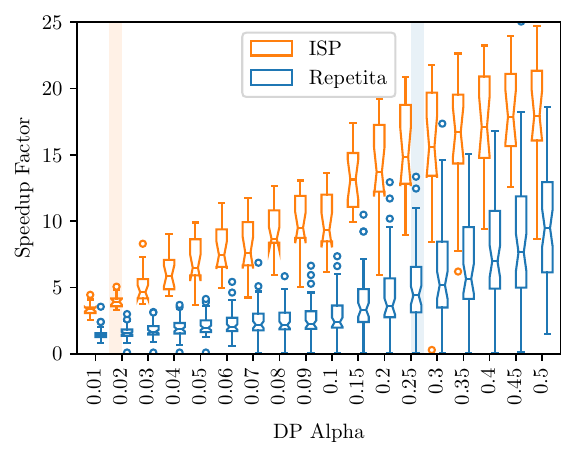}
         \caption{Demand Pinning}
         \label{subfig:demand_pinning_speedup}
     \end{subfigure}
        \caption{Achievable speedup for the 2SR optimization. (A few very large outliers were cut off for better readability.)}
        \label{fig:approach_comparison_speedup}
\end{figure*}

In this section, we evaluate the performance of the various middlepoint selection approaches presented in Section \ref{sec:sr_middlepoint_selection}.
The \ac{MLU} deteriorations and the achievable speedup for different parameterizations are depicted in Figure \ref{fig:approach_comparison_mlu} and \ref{fig:approach_comparison_speedup}, respectively, with individual subfigures for each approach.
Orange boxplots show the respective distributions for the Tier-1 ISP dataset and blue boxplots for the Repetita dataset.

\subsection{A Primer Regarding CPLEX-Related Outliers}
In rare occasions, there can be outliers with a speedup factor below one (e.g., in Figure \ref{subfig:demand_pinning_speedup} for $\alpha_{DP} = 0.3$ for the ISP data).
This means using the respective preprocessing approach actually resulted in an increase in computation time.
While this is rather surprising at first thought, there is a rather simple explanation for this phenomenon. 
\ac{LP}-solvers like CPLEX stop optimization only if they find a ``proof'' that the currently best solution is truly optimal (or within a small margin to the optimum).
This \textit{optimality gap} is computed by comparing the currently best found solution against a lower bound for the best possible objective value which is continuously updated (increased) during optimization.
If the gap between the lower bound and the best found solution is sufficiently small, the solution is considered to be optimal.
By preemptively limiting the allowed set of available \ac{SR} paths, the rare scenario can occur in which we prohibit a path that might not be required for an optimal solution but that facilitates a quick proof of optimality.
There might be other options for such a proof but if these are explored in a much later stage of the \textit{branch-and-cut search}, proving optimality and, thus, the whole optimization process might take substantially longer.
The same effect is also responsible for the more noticeable ``drop'' of the achievable speedup when going from 10 to a single digit number of middlepoints for the \ac{ISP} dataset in Figure \ref{subfig:centrality_speedup}.
\par
\par
While the possibility of actually degrading performance when applying preprocessing approaches might be concerning, there is a straight-forward solution.
Increasing the allowed \textit{optimality gap} of CPLEX by only a small amount allows for an easier proof of ``optimality'' (even without the paths excluded by the preprocessing).
In our experiments, increasing the optimality gap from the default $10^{-4}$ to around $10^{-3}$ proved promising to resolve these issues without having a practically relevant negative impact on the solution quality.

\subsection{Centrality-based Middlepoint Selection}
For the centrality-based middlepoint selection (Figures \ref{subfig:centrality_mlu} and \ref{subfig:centrality_speedup}), it can be seen that allowing only a small set of nodes as available middlepoints can result in a substantial (factor 10--20) speed-up in computation time.
This, however, comes at the price of a significant deterioration of the overall solution quality.
In the worst cases, \acp{MLU} increase by up to 55\% for the \ac{ISP} dataset and by more than 80\% for the Repetita instances.
By increasing the number of allowed middlepoints, these \ac{MLU} deteriorations can be reduced but this also results in an increase in computation time.
Ultimately, operators have to make an individual decision regarding the acceptable trade-off between speedup and resulting \ac{MLU} deterioration.
This might vary for different use cases, but from our experience, the highest acceptable \ac{MLU} deterioration for most scenarios probably lies somewhere around 10-15\%, at most.
Based on this, we highlight in each plot the parameter configuration that produces the highest speedup while still allowing for, in our experience, practically usable \acp{MLU}.
For the Repetita data this is around 25 middlepoints.
For the ISP instances,  the box and whiskers are already below the 10\% deterioration threshold for just three middlepoints.
However, since the dataset only comprises 18 instances, the three ``outliers'' that (significantly) surpass this threshold still make up over 15\% of the dataset.
To improve the solution quality for these instances, substantially higher numbers of middlepoints are needed (in the range of 25-45).
Hence, we argue that the number of middlepoints required to obtain practically usable solutions is (to some extend) instance-dependent but, in general, at least 20 to 25 middlepoints seem to be required.
This translates to an average speedup factor of around 3-4 for the Repetita dataset and 7-8 for the \ac{ISP} backbone (cf. Figure \ref{subfig:centrality_speedup}).
It has to be noted, however, that for these parameterizations there is still a considerable number of instances with a significant \ac{MLU} deterioration left.
Overall, our results generally confirm the findings of \cite{centrality_based_middlepoint_selection}.
Selecting only a few central nodes as available middlepoints can substantially reduce computation times but also results in significant deteriorations of the overall \acp{MLU}, especially for small numbers of middlepoints.

\subsection{Stretch-Bounding}
In Figure \ref{subfig:stretchbounding_mlu} and \ref{subfig:stretchbounding_speedup}, it can be seen that, for the \ac{ISP} dataset, near optimal results are achieved with a \ac{SB} factor of just $1.05$.
For a factor of $1.1$, there is basically no noticeable \ac{MLU} deterioration anymore while still achieving a speedup of factor 4 to 5.
For the Repetita dataset, the results differ noticeably.
Here, such low \ac{SB} factors result in a substantial \ac{MLU} deterioration of around 40\% on average and over 80\% at max.
Practically usable results can be achieved with a \ac{SB} factor of $1.5$ or higher and (virtual) optimal results require a factor of around $2.0$.
This translates to a speedup of around factor 4.5 and factor two, respectively.
\par
At first glance, it might seem like the \ac{SB} approach performs substantially worse for the Repetita dataset.
This observation, however, is (at least a bit) deceptive.
While, for low \ac{SB} factors, the \ac{MLU} deterioration on the Repetita dataset is substantially higher, the speedup is also much better.
The reason for this is that for the same \ac{SB} factor, the overall number of prohibited \ac{SR} paths is much higher for the Repetita dataset.
For example, for a \ac{SB} factor of $1.1$, around 90\% or more of all available \ac{SR} paths are prohibited for many Repetita instances.
Contrary, for the \ac{ISP} data, only around 65-70\% of paths are filtered out.
As a result, the optimization for the Repetita instances is faster due to the lower number of options to evaluate, but this also results in a worse overall solution quality.
If we instead compare results based on the percentage of excluded \ac{SR} paths, they become much more similar.
For example, for a \ac{SB} factor of $1.4$, the percentage of excluded \ac{SR} paths is also in the range of 65-70\% and the resulting speedup is comparable to the one of the \ac{ISP} data with the respective ``matching'' \ac{SB} factor of $1.1$.
We suspect that these differences in the number of excluded \ac{SR} paths are a product of topological differences between the instances in the Repetita dataset and the real \ac{ISP} backbone network.
However, investigating and identifying these differences is out of the scope of this work, but remains an interesting question for future work.

\subsection{Demand Pinning}
Results for the \ac{DP} approach are depicted in Figures \ref{subfig:deman_pinning_mlu} and \ref{subfig:demand_pinning_speedup}.
It can be seen that for the \ac{ISP} data, a speedup of around factor $4.5$ can be achieved without substantially worsening the \ac{MLU}.
However, for larger $\alpha$-values, the solution quality deteriorates quickly.
Results for the Repetita data look rather different.
Here, we are able to exclude up to 20\% and more of the total traffic volume before a relevant \ac{MLU} deterioration becomes observable.
However, the speedup, while overall slightly better than for the \ac{ISP} data, remains rather similar with a factor of around 5.
The reason for those differences lies in the distribution of the demand sizes in the traffic matrices of the two datasets.
The real \ac{ISP} traffic features a significantly higher number of really small demands (w.r.t. the total traffic) than the artificially generated matrices in the Repetita dataset.
This is exemplarily depicted in Figure \ref{fig:demand_size_comparison} for the largest instance of the \ac{ISP} and Repetita dataset, respectively.
As a result, the same $\alpha$-value allows for the exclusion of substantially more demands for the \ac{ISP} network.
An $\alpha$-value of $0.01$, for example, allows to exclude around 70-80\% of all demands in the \ac{ISP} matrices from optimization, while only excluding around 15-20\% of demands from the Repetita matrices.

\subsection{SR Path Domination}
\begin{figure}[t]
\centering
\begin{minipage}[t]{0.475\linewidth}
\centering
\includegraphics[width=0.95\linewidth]{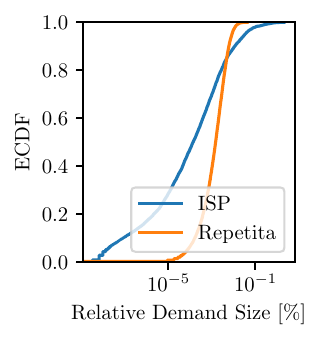}
\caption{ECDF of the relative demand sizes (w.r.t to the total traffic volume) of the \ac{ISP} \texttt{2021} and the Repetita \texttt{Cogentco} instances.}
\label{fig:demand_size_comparison}
\end{minipage}%
\hfill%
\begin{minipage}[t]{0.475\linewidth}
\centering
\includegraphics[width=0.95\linewidth]{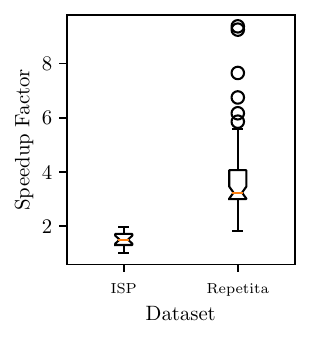}%
\caption{2SR speedup achieved by the \textit{\ac{SR} path domination} preprocessing on the two evaluation datasets.}%
\label{fig:pd_speedup}
\end{minipage}
\end{figure}

Figure \ref{fig:pd_speedup} depicts the speedup that is achievable with the \ac{SR} path domination approach on our two datasets.
Contrary to the previous approaches, there is no need to look at \ac{MLU} deterioration since the main idea of the \ac{SR} path domination concept is to retain provable optimality of the achievable \acp{MLU}.
It can be seen that, for the Repetita data, computation times can be improved by around a factor of four on most instances with a couple of outliers even reaching close to factor 10.
Those high outliers are a result of the special topology structures of certain instances. For example, the \texttt{Ulaknet} topology consists three star shaped networks who's centers are interconnected with each other.
Basically all \ac{SR} paths using one of the many stub-nodes as intermediate segment provide no \ac{TE} benefit regarding the \ac{MLU} and can be ignored. This results in over 98\% of all \ac{SR} paths being ignored for optimization. 
On more ``realistically'' shaped topologies, however, this number is much lower (mostly between 65-80\%) and, hence, the achievable speedup is also more moderate.
\par
While the \ac{SR} path domination preprocessing works quite well for the Repetita data, this does not hold for the real-world \ac{ISP} network.
Here, the average achievable speedup factor is just around 1.5 and the maximum barely surpasses factor 2.
The reason for this, again, lies in the number of \ac{SR} paths that are ruled out for each respective dataset.
For the \ac{ISP} dataset, this number is substantially lower with just around 20\% of the total number of \ac{SR} paths compared to an average of around 80\% for most Repetita instances.
We do not have a definitive answer what causes this behavior but we suspect that it is a result of topological differences between the networks in the Repetita dataset and the real \ac{ISP} network.
For example, the \ac{ISP} network has virtually no stub-nodes since a common design goal for modern networks is to achieve at least two-connectivity for all nodes.
This facilitates reliability and robustness as it ensures that the network will not be partitioned by single-link failures.
Contrary, the Repetita topologies feature a rather large number of stub-nodes.
Since those are never needed as middlepoints to obtain an optimal solution, a larger number of stub nodes automatically results in a larger number of dominated \ac{SR} paths.
This becomes visible when removing all stub-nodes from the Repetita topologies which reduces the number of excluded \ac{SR} paths from around 80\% on average to just 60\%.
For reasons of space, we cannot delve deeper into this topic here and leave it for future work.

\subsection{Discussion}
All in all, we have seen that each preprocessing approach has its pros and cons.
Some perform better on the Repetita data and some on the \ac{ISP} data.
Hence, there is no clear ``winner'' to be picked but we hope that our extensive analysis facilitates others in picking a suitable preprocessing approach for their applications.
However, what became more and more clear during our evaluation, is the fact that considerable differences in results are observable depending on the dataset used.
This especially holds true for the \ac{SR} path domination approach that works really well for Repetita networks but significantly worse for the real Tier-1 \ac{ISP} backbone.
We track this down to differences in topology and traffic characteristics between the two datasets.
This reinforces our concerns regarding the direct transferability of results obtained on the Repetita data with artificial traffic to real networks, already expressed in Section \ref{subsec:rel_work_discussion}.
It also stresses the importance of also carrying out evaluations on real, recent network data.
Of course, our \ac{ISP} dataset is ``just one datapoint'' that does not allow to draw definitive and universal conclusions but other recently reported information e.g., regarding the demand size distribution in the \textit{Microsoft} network \cite{demand_pinning_1} is far closer to the \ac{ISP} network characteristics than to the Repetita data.

\begin{figure*}
     \centering
     \begin{subfigure}[b]{0.32\textwidth}
         \centering
         \includegraphics[width=\textwidth]{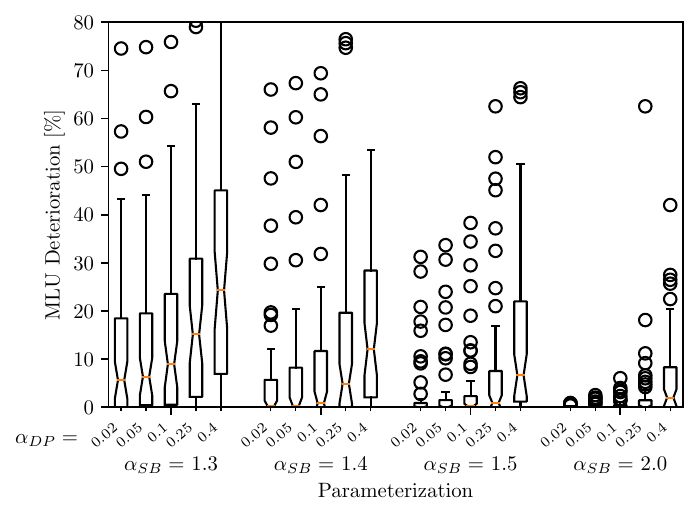}
         \caption{2SR on Repetita dataset}
         \label{subfig:combined_repetita_mlu}
     \end{subfigure}
     \hfill
     \begin{subfigure}[b]{0.32\textwidth}
         \centering
         \includegraphics[width=\textwidth]{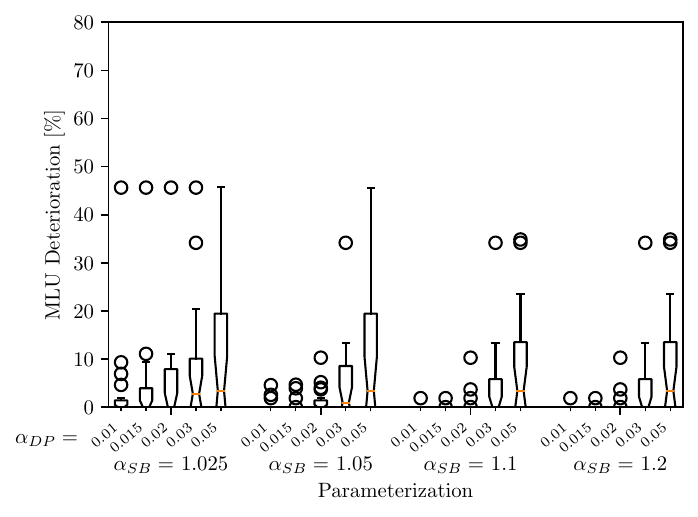}
         \caption{2SR on ISP dataset}
         \label{subfig:combined_isp_mlu}
     \end{subfigure}
     \hfill
     \begin{subfigure}[b]{0.32\textwidth}
         \centering
         \includegraphics[width=\textwidth]{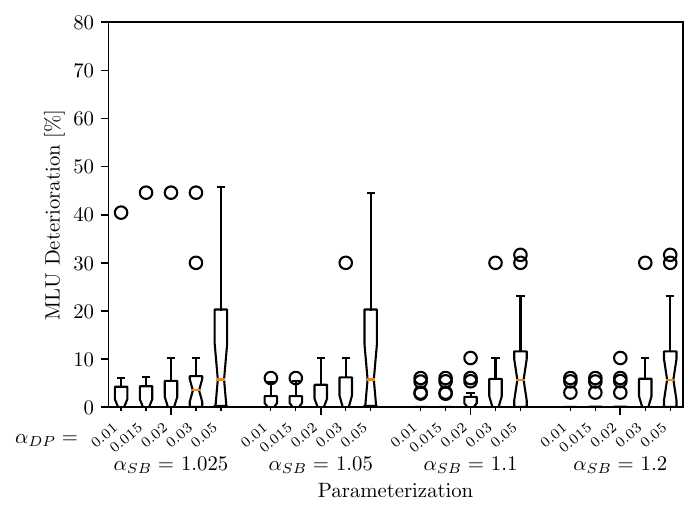}
         \caption{SC2SR on ISP dataset}
         \label{subfig:combined_isp_mo_mlu}
     \end{subfigure}
        \caption{MLU deterioration resulting from the combined preprocessing approach for different datasets and \ac{SR} algorithms.}
        \label{fig:combined_preprocessing_mlu_deterioration}
\end{figure*}

\begin{figure*}
     \centering
     \begin{subfigure}[b]{0.32\textwidth}
         \centering
         \includegraphics[width=\textwidth]{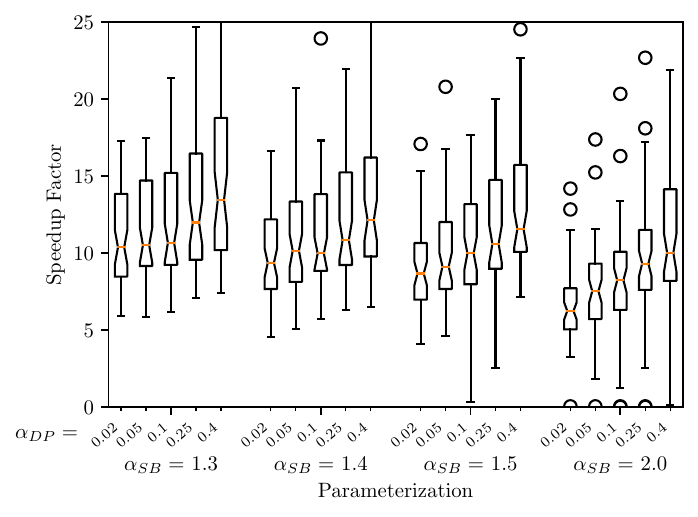}
         \caption{2SR on Repetita dataset}
         \label{subfig:combined_repetita_speedup}
     \end{subfigure}
     \hfill
     \begin{subfigure}[b]{0.32\textwidth}
         \centering
         \includegraphics[width=\textwidth]{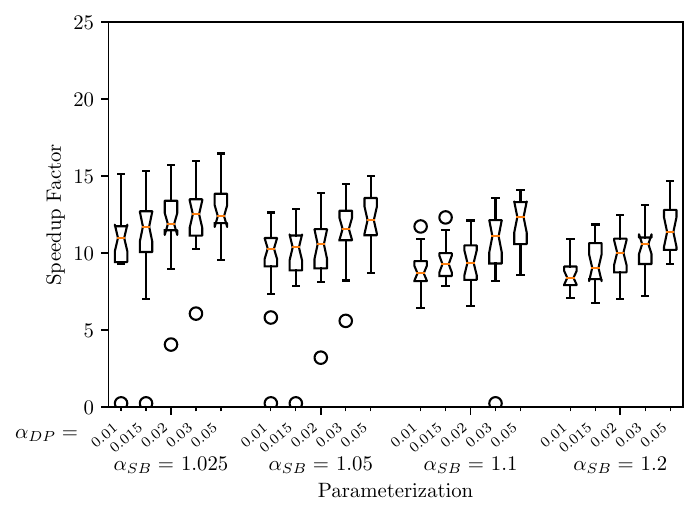}
         \caption{2SR on ISP dataset}
         \label{subfig:combined_isp_speedup}
     \end{subfigure}
     \hfill
     \begin{subfigure}[b]{0.32\textwidth}
         \centering
         \includegraphics[width=\textwidth]{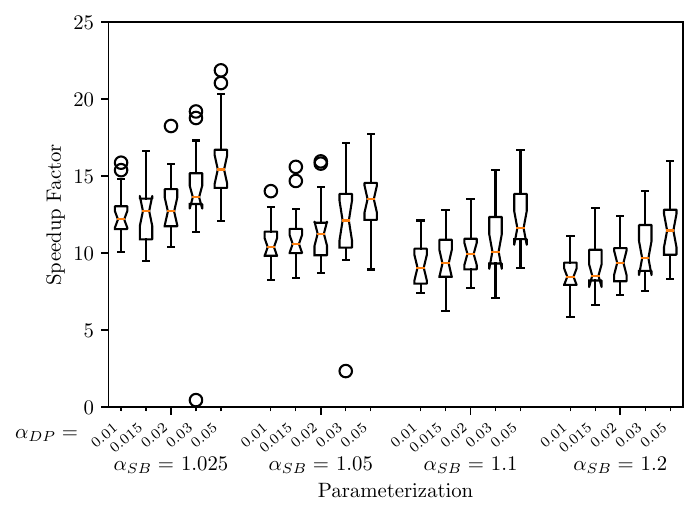}
         \caption{SC2SR on ISP dataset}
         \label{subfig:combined_isp_mo_speedup}
     \end{subfigure}
        \caption{Speedup achieved by the combined preprocessing approach for different datasets and \ac{SR} algorithms.}
        \label{fig:combined_preprocessing_speedup}
\end{figure*}

\section{Combining Preprocessing Approaches to Improve Performance}
As seen before, there is no definitive answer to what the generally best preprocessing approach is since performance varies between datasets.
In this section we propose the concept of combining multiple preprocessing approaches to allow for a more consistent performance across all datasets and to further improve the achievable speedup.

\subsection{Concept}
Our approach is based on the observation that, until now, we always considered each preprocessing approach individually.
However, they are not mutually exclusive.
Therefore, it is possible to combine multiple preprocessing approaches into a single one.
This can yield multiple benefits.
First and foremost, it yields the potential of further increasing the achievable speedup.
Combining the individual sets of excluded \ac{SR} paths allows to further increase the number of \ac{SR} paths that can be ignored during optimization.
However, it is unclear whether the combination of multiple exclusion sets that perform well individually will result in a well performing union set, as well.
The combined set might also become too restrictive and, hence, might result in substantial \ac{MLU} deterioration.
Secondly, combining different preprocessing approaches might also lead to a more ``stable'' performance across our two datasets.
In simple terms, by combining an approach that works better on the \ac{ISP} data (i.e., \ac{SB}) with one that seems more suited for the Repetita data (i.e., \ac{SR} path domination), we hope to leverage their individual benefits and get an algorithm that performs well on both datasets.
\par
For our improved preprocessing algorithms, we combine the three approaches of \ac{SB}, \ac{DP} and \ac{SR} path domination.
The reason for not including the centrality-based approach is that it generally performs worse than the other approaches with regards to \ac{MLU} deterioration and speedup.
Furthermore, as already discussed in Section \ref{subsec:centrality_based}, it also features other weaknesses when it comes to practical use (e.g., regarding latency constraints).
Our new combined preprocessing approach starts with a \ac{DP} operation that can be fine-tuned with the $\alpha_{DP}$ parameter.
After this, a \ac{SB} step is carried out using the $\alpha_{SB}$ parameter.
The \ac{SR} path domination filtering comes last as it is the computationally most demanding operation. Having already filtered out a large set of \ac{SR} paths by the previous two operations which do not need to be checked for domination or equivalency anymore, facilitates lower computation times. 
\par

\subsection{Evaluation}
We evaluate the performance of our combined approach on our two evaluation datasets for the 2\ac{SR} algorithm.
Additionally, we also carry out a short exemplary evaluation for the \ac{MO}-capable SC2SR algorithm proposed in \cite{midpoint_optimization_infocom}.
This is done to provide an insight into whether results are transferable to other \ac{SR} \ac{TE} algorithms even if those utilize a rather different \ac{SR} variation.
The results regarding \ac{MLU} deterioration and speedup for various parameter combinations are depicted in Figures \ref{fig:combined_preprocessing_mlu_deterioration} and \ref{fig:combined_preprocessing_speedup}, respectively.
It can be seen that for the Repetita dataset (Figures \ref{subfig:combined_repetita_mlu} and \ref{subfig:combined_repetita_speedup}), with the right parameter configuration, we are able to achieve a 2SR speedup of around factor 7 to 8 without a significant \ac{MLU} deterioration. If a few more outliers are acceptable, this factor can be increased even further close to a ten-times speedup.
\begin{table}
	\centering
	\caption{Computation time information (in seconds) for the respective ground truth algorithm without preprocessing.}
	\label{tab:comp_times}
	\small
	\begin{tabular}{l c c c c c}
		\toprule
		&& Min & Max & Median & Average\\
		\midrule
		\multirow{2}{*}{2SR} & Repetita & 10 & 5456 & 37 & 311 \\
		& ISP & 177 & 1401 & 705 & 630 \\
		SC2SR & ISP & 2477 & 14568 & 5968 & 6670 \\
		\bottomrule
	\end{tabular}
\end{table}
The same applies to the \ac{ISP} dataset for both the 2SR (Figures \ref{subfig:combined_isp_mlu} and \ref{subfig:combined_isp_speedup}) and the \ac{MO}-capable SC2SR algorithm (Figures \ref{subfig:combined_isp_mo_mlu} and \ref{subfig:combined_isp_mo_speedup}).
This confirms that the new preprocessing approach performs (more or less) equally good across both datasets when the right parameters are chosen.
Furthermore, it also shows that the preprocessing is transferable to other \ac{SR} optimization algorithms.
It achieves comparable results regarding \ac{MLU} deterioration and speedup, even if the underlying \ac{SR} concept is inherently different.
To put it into perspective what a speedup of around factor 10 actually means, some information on the computation times of the ground-truth algorithms (without any preprocessing) are given in Table \ref{tab:comp_times}.
It can be seen that, for example, the SC2SR algorithm takes around two hours to compute, on average, and over four hours at max. With our preprocessing, this can be reduced to just around 10 or 20 minutes, respectively.
The benefits of our preprocessing become even more apparent when looking at the 2SR algorithm.
Here, our preprocessing is able to reduce the average computation time from 10min and more, to less than 2min for most of the \ac{ISP} instances.
This easily allows for the use of \ac{LP}-based optimization for use cases where network configuration is continuously adapted on a timescale of just a few minutes (cf. e.g., \cite{microsoft_swan}).
Furthermore, we are now advancing into computation time regions in which it can be argued that the 2SR algorithm could even be used for tactical \ac{TE} that allows to quickly react to failures or traffic shifts \cite{mols_heuristic}.
Finally, we also believe that the performance achieved with our preprocessing approach is reasonably close to what can actually be achieved with preprocessing in general.
The reason for this are its extremely high numbers of excluded \ac{SR} paths.
For the 2SR algorithm, for example, our preprocessing already rules out 97-99\% of all theoretically configurable 2SR paths.
This probably does not leave much room for further improvement since a certain number of options to choose from is required before solution quality starts to degrade substantially.

\section{Conclusion} \label{sec:conclusion}
In this paper, we conducted the first large scale comparative study of existing preprocessing (or \textit{middlepoint selection}) approaches for \ac{SR}.
For this, we not only used publicly available data from the \textit{Repetita} dataset, but also real network data from a globally operating Tier-1 \ac{ISP}.
Based on the insights gained from this study, we propose a combination of multiple preprocessing approaches to further improve performance.
We are able to show that, with this approach, computation times of different \ac{LP}-based optimization algorithms can be reduced by a factor of 10 or more without a significant deterioration in solution quality.
This represents a major improvement over the current state-of-the-art and facilitates the reliable usability of \ac{LP}-based \ac{SR} \ac{TE} in large networks.

\newpage

\balance

\bibliographystyle{IEEEtranS}
\bibliography{IEEEabrv,main}

\end{document}